\documentclass[conference, letterpaper, 10pt]{IEEEtran}
\usepackage[utf8]{inputenc}
\IEEEoverridecommandlockouts
\usepackage{amsmath,amssymb,amsfonts}
\usepackage{graphicx}
\usepackage{textcomp}
\usepackage{xcolor}
\usepackage[pdftex,bookmarks=false]{hyperref}

\usepackage[ruled, vlined, linesnumbered, longend, resetcount]{algorithm2e}

\SetCommentSty{mycommfont}

\usepackage{amsmath}
\usepackage{mathtools}
\usepackage{tabularx}
\usepackage{amsfonts}
\usepackage{booktabs}
\usepackage{algpseudocode}

\usepackage{graphicx}
\usepackage{lipsum}
\usepackage{float}
\usepackage{afterpage}
\usepackage{adjustbox}

\usepackage{orcidlink}
\usepackage{authblk}
\usepackage{placeins}

\usepackage[backend=biber, style=ieee]{biblatex}
\renewbibmacro{in:}{\ifentrytype{article}{}{\printtext{\bibstring{in}\intitlepunct}}}

\begin{document}

\title{Distributed Q-learning-based Shortest-Path Tree Construction in IoT Sensor Networks\\

\thanks{This work was partly supported by the Korea government (MSIT), IITP, Korea, under the ICT Creative Consilience program (IITP-2025-RS-2020-II201821, 20\%); AI Innovation Hub (RS-2021-II212068, 20\%), the Development of 6G Network Integrated Intelligence Plane Technologies (IITP-2025-RS-2024-00392332, 30\%); and the National Research Foundation of Korea (RS-2024-00343255, 30\%).}
}

\author[1]{Van-Vi Vo  \orcidlink{0000-0002-5745-4164}}
\author[2]{Tien-Dung Nguyen \orcidlink{0000-0003-0064-4044}}
\author[3,*]{Duc-Tai Le \orcidlink{0000-0002-5286-6629}}
\author[3,*]{Hyunseung Choo \orcidlink{0000-0002-6485-3155}}
\affil[1]{{\small Convergence Research Institute, Sungkyunkwan University, Suwon, South Korea}}
\affil[2]{{\small School of Electrical and Electronic Engineering, Hanoi University of Science and Technology, Hanoi, Vietnam}}
\affil[3]{{\small Dept. of Electrical and Computer Engineering, Sungkyunkwan University, Suwon, South Korea}}
\affil[*]{{\small Corresponding authors (\{ldtai, choo\}@skku.edu)}}


\maketitle

\begin{abstract}
Efficient routing in IoT sensor networks is critical for minimizing energy consumption and latency. Traditional centralized algorithms, such as Dijkstra's, are computationally intensive and ill-suited for dynamic, distributed IoT environments. We propose a novel distributed Q-learning framework for constructing shortest-path trees (SPTs), enabling sensor nodes to independently learn optimal next-hop decisions using only local information. States are defined based on node positions and routing history, with a reward function that incentivizes progression toward the sink while penalizing inefficient paths. Trained on diverse network topologies, the framework generalizes effectively to unseen networks. Simulations across 100 to 500 nodes demonstrate near-optimal routing accuracy (over 99\% for $N \geq 300$), with minor deviations (1-2 extra hops) in smaller networks having negligible impact on performance. Compared to centralized and flooding-based methods, our approach reduces communication overhead, adapts to topology changes, and enhances scalability and energy efficiency. This work underscores the potential of Q-learning for autonomous, robust routing in resource-constrained IoT networks, offering a scalable alternative to traditional protocols.

\end{abstract}


\section{Introduction}

The rapid growth of Internet of Things (IoT) sensor networks has transformed applications such as environmental monitoring, smart cities, and industrial automation, relying on low-power sensor nodes to collect and transmit data to a central sink via a convergecast model. A critical challenge in these networks is constructing a shortest-path tree (SPT), where each node's path to the sink minimizes hop count or other metrics like latency and energy consumption. Efficient SPT construction reduces communication overhead, conserves energy, and ensures low-latency data delivery, which are paramount in resource-constrained IoT environments. However, achieving optimal routing in large-scale, distributed IoT networks is complex due to limited computational and energy resources, decentralized operation, and dynamic network conditions such as node failures or link variations.

Traditional routing algorithms, such as Dijkstra’s \cite{dijkstra2022note} and Bellman-Ford \cite{bellman1958routing}, compute optimal paths using global network topology knowledge, necessitating energy-intensive exchanges of state information that are impractical for IoT nodes with constrained battery life and introducing single points of failure \cite{tanenbaum2011}. Distributed protocols like distance-vector or link-state routing mitigate some issues but require continuous topology updates, leading to significant communication overhead. Greedy geographic routing, such as GPSR \cite{karp2000}, forwards packets to neighbors closer to the sink but often fails in the presence of local optima, resulting in suboptimal or failed deliveries. Flooding-based protocols, like RPL \cite{winter2012}, construct breadth-first search trees but suffer from the broadcast storm problem, causing redundant transmissions and excessive energy consumption \cite{tseng2002}. Recent surveys highlight that probabilistic methods, such as gossip-based routing, reduce overhead but sacrifice path optimality and scalability in dense networks \cite{gaidhani2023review}.

Reinforcement learning (RL) has recently gained traction for adaptive routing in wireless networks. Q-routing \cite{boyan1994}, a pioneering RL approach, enables nodes to learn optimal paths by updating Q-values based on routing outcomes, offering adaptability to dynamic conditions. Recent advancements, such as RL-based routing for energy efficiency \cite{yuan2024deep}, near-tight shortest paths in hybrid models \cite{schneider2023neartight}, and graph convolutional network-based SPT construction \cite{park2024efficient}, demonstrate improved performance over traditional methods. However, most RL approaches either rely on centralized training, require extensive communication for coordination, or lack formal guarantees for distributed SPT construction in IoT contexts. Moreover, they often do not address the unique constraints of IoT sensor networks, such as ultra-low power budgets and the need for loop-free routing trees.

To bridge this gap, we propose a distributed Q-learning framework tailored for SPT construction in IoT sensor networks. Our approach enables each node to independently learn optimal next-hop decisions using only local information, eliminating centralized control and minimizing communication overhead. Our key contributions are summarized as follows:
\begin{itemize}
    \item We introduce a fully distributed Q-learning algorithm allowing each node to independently and collaboratively learn optimal next-hop decisions, requiring only localized information exchange.

    \item We formally define the network model and problem, with theoretical guarantees for method convergence and optimality. We also analyze computational and communication complexity, showing significant reductions over traditional approaches.

    \item Extensive simulations confirm that our approach achieves superior performance in terms of convergence speed, optimal path discovery, energy efficiency, and adaptability to network dynamics, outperforming existing routing algorithms.
\end{itemize}

The paper is structured as follows: Section II defines the network model and problem. Section III presents the Q-learning methodology. Section IV details experimental results. Finally, we conclude our work with future research directions in Section V.

\section{Network Model and Problem Statement}
\subsection{Network Topology and Assumptions}

We model an IoT sensor network as an undirected graph $G = (V, E)$, where $V$ is the set of sensor nodes with $|V| = N$, and $E$ is the set of wireless communication links. A link $(u, v) \in E$ exists if nodes $u$ and $v$ are within the communication range and have sufficient link quality for reliable data exchange. We assume the network is connected, ensuring at least one multi-hop path between any pair of nodes. A designated sink node $v_0 \in V$ serves as the central data collection point, typically a gateway with enhanced computational resources or external power supply, acting as the root of the routing topology. Nodes generate data periodically or event-driven, requiring efficient forwarding to $v_0$.

Each link $(u, v) \in E$ has an associated cost $w(u, v)$, which may represent hop count, distance, latency, or transmission energy. For clarity, we focus on hop count with $w(u, v) = 1$ per link, though the methodology extends to other non-negative additive metrics. We assume symmetric links, i.e., $w(u, v) = w(v, u)$, typical in low-power IoT deployments with symmetric radios and balanced antennas, simplifying routing. Wireless communication faces constraints like limited transmission range, signal interference, and collisions. We assume an idealized MAC layer ensuring reliable packet delivery via retries.

\subsection{Shortest-Path Tree (SPT) Problem Formulation}

Given graph $G = (V, E)$ and sink $v_0$, the goal is to construct an SPT $T = (V, E_T)$ rooted at $v_0$, with $E_T \subseteq E$. Tree $T$ spans all nodes in $V$, each (except $v_0$) having one parent, satisfying:
\begin{equation}
    d_T(v, v_0) = d_G(v, v_0), \quad \forall v \in V,
\end{equation}
where $d_T(v, v_0)$ is the path length in $T$, and $d_G(v, v_0)$ is the shortest path in $G$. This minimizes costs and delays. Multiple SPTs may exist; our distributed algorithm lets nodes select optimal parents, forming a global SPT.

SPT construction follows IoT constraints:
\begin{itemize}
    \item \textbf{Decentralized Operation}: No centralized controller or global knowledge, except the sink's root role, for scalability and robustness.
    \item \textbf{Local Information}: Nodes use only one-hop neighbor data, minimizing non-local communication.
    \item \textbf{Efficient Communication}: Small control messages for route establishment and maintenance, conserving resources.
    \item \textbf{Stability}: Algorithm stabilizes to sustainable topology with minimal overhead.
\end{itemize}

Loop-freedom is critical, as the tree must be acyclic to avoid circulating packets. Ensuring this during and after learning is a key challenge for our Q-learning approach. Key notations are summarized in Table~\ref{tab:notation}.

\begin{table}[htbp]
\centering
\caption{Summary of Notations}
\resizebox{\columnwidth}{!}{
\begin{tabular}{|c|l|}
\hline
\textbf{Notation} & \textbf{Description} \\ \hline
$v_0$ & Sink node (root of the SPT). \\ \hline
$G = (V, E)$ & Network graph with nodes $V$ and edges $E$. \\ \hline
$N(v)$ & Set of one-hop neighbors of node $v$. \\ \hline
$w(u, v)$ & Weight of link $(u, v)$, typically 1 (hop count). \\ \hline
$d_G(v, v_0)$ & Shortest path length from $v$ to $v_0$ in $G$. \\ \hline
$d_T(v, v_0)$ & Path length from $v$ to $v_0$ in tree $T$. \\ \hline
$Q(v, u)$ & Q-value for node $v$ selecting neighbor $u$. \\ \hline
$\pi(v)$ & Parent node (next hop) of $v$ in $T$. \\ \hline
$\alpha$ & Learning rate for Q-value updates. \\ \hline
$\gamma$ & Discount factor for future rewards. \\ \hline
$\epsilon$ & Exploration probability in $\epsilon$-greedy policy. \\ \hline
\end{tabular}}
\label{tab:notation}
\end{table}

\section{Proposed Approach}
To address the SPT construction problem in IoT sensor networks, we propose a distributed Q-learning framework that enables each sensor node to independently learn optimal next-hop decisions using only local information. This reinforcement learning approach eliminates the need for centralized control or global network knowledge, making it suitable for resource-constrained IoT environments. \textbf{Note:} Throughout this section, \emph{nodes} refer to their \emph{locations} — i.e., node $v$ denotes location $(x_v, y_v)$, not a node ID. Below, we describe the Q-learning model components, the $\epsilon$-greedy exploration strategy, the Q-value update mechanism, and the distributed algorithm for both training and testing phases.

\subsection{Q-Learning Model Components}

The IoT sensor network is modeled as an undirected graph $G = (V, E)$, where $V$ is the set of sensor nodes at locations $(x_v, y_v)$ ($|V| = N$), $E$ represents bidirectional wireless communication links with Euclidean distance $\leq R$, and the sink node $v_0$ has a fixed location $(x_0, y_0) \in V$, typically at the center of a square deployment area of size $W \times W$. Let $N(v) = \{ u \in P \mid \sqrt{(x_u - x_v)^2 + (y_u - y_v)^2} \leq R \}$ be the set of \textbf{grid neighbors} of $v$ in the full $W \times W$ area, where $P = [0,W) \times [0,W)$. For a specific graph $G_m = (V_m, E_m)$, define $N^{V_m}(v) = N(v) \cap V_m$ as the \textbf{graph-specific neighbors}. In the Q-learning framework, we define the following components:

\begin{itemize}
    \item \textbf{State}: The current node $v \in V$ where the agent (representing a packet or routing decision) resides.
    \item \textbf{Action}: Selection of a next-hop node $u \in N^{V_m}(v)$.
    \item \textbf{Reward}: The immediate reward $r(v,u)$ for selecting action $u$ from node $v$ is:
    \begin{equation}
        r(v,u) =
        \begin{cases}
            100 & \text{if } u = v_0, \\
            0 & \text{if } u \in N^{V_m}(v) \setminus \{v_0\}.
        \end{cases}
        \label{Eq: reward}
    \end{equation}
    This reward structure incentivizes shortest paths by assigning a high positive reward only when the sink is reached.
    \item \textbf{Episode}: Starts at a random node $v \in V_m \setminus \{v_0\}$ and proceeds by selecting next-hop nodes in $N^{V_m}(v)$ until $v_0$ is reached or a termination condition (e.g., no valid neighbors) is met.
\end{itemize}

Each node $v \in V_m \setminus \{v_0\}$ maintains a Q-table with entries $Q(v, u)$ for each neighbor $u \in N(v)$, estimating the expected cumulative reward of selecting $u$ as the next hop. After convergence, the optimal policy $\pi(v) = \arg\max\limits_{u \in N(v)} Q(v, u)$ defines the parent node for $v$ in the SPT.


\subsection{Exploration and Update Mechanisms}

To balance exploration and exploitation, we employ an $\epsilon$-greedy policy. At node $v$ in graph $G_m = (V_m, E_m)$, the next-hop $u$ is selected from the \textbf{graph-specific neighbor set} $N^{V_m}(v) = \{ u \in V_m \mid u \in N(v) \}$ as follows:
\begin{itemize}
    \item With probability $\epsilon$: $u \sim \text{Uniform}(N^{V_m}(v))$ \quad (exploration),
    \item With probability $1-\epsilon$: $u = \arg\max\limits_{u' \in N^{V_m}(v)} Q(v, u')$ \quad (exploitation, random tie-breaking).
\end{itemize}
After selecting action $u$, observing reward $r = R(v, u)$, and transitioning to $u$, the Q-value is updated using the Bellman equation (Equation~\ref{eq:bellman}):
\begin{equation}
    Q(v, u) \leftarrow (1 - \alpha) Q(v, u) + \alpha \left( r(v,u) + \gamma \max\limits_{u' \in N^{V_m}(u)} Q(u, u') \right),
    \label{eq:bellman}
\end{equation}
where $\alpha \in (0,1]$ is the learning rate and $\gamma \in [0,1)$ is the discount factor. If $N^{V_m}(u) = \emptyset$, the max term is $0$. This update ensures convergence to optimal routing under sufficient exploration.

\subsection{Distributed Q-Learning Algorithm}

The Q-learning algorithm operates in two phases: a \emph{training phase} to learn the Q-table across multiple network instances and a \emph{testing phase} to construct SPTs on unseen networks using the trained Q-table. Each node $v \in V_m \setminus \{v_0\}$ runs the algorithm asynchronously, exchanging only maximum Q-value summaries with neighbors in $N^{V_m}(v)$, minimizing communication overhead. The training phase is detailed in Algorithm~\ref{alg:qlearning_training}; the testing phase in Algorithm~\ref{alg:qlearning_testing}.

\begin{algorithm}[htbp]
\small
\caption{Q-Learning Training for SPT Construction}
\label{alg:qlearning_training}
\KwIn{Sink $v_0$ (fixed location), communication range $\mathcal{R}$, learning rate $\alpha$, discount factor $\gamma$, exploration probability $\epsilon$, number of training graphs $M$, episodes per graph $K$, square grid size $W \times W$}
\KwOut{Trained Q-matrix}


\tcp{Nodes mentioned in the algorithm refer to their locations $(x, y)$. $N(v)$: neighbors of $v$ in $|W|\times|W|$ ($\leq \mathcal{R}$). $N^{V}(v)$: neighbors of node $v$ in graph $G=(V,E)$}

\tcp{Set of locations in a grid area $|P| = W^2$}
Create $P=\{(x,y)\}\big| x,y \in [0,W)$\;

\tcp{Q-matrix init}
$Q \in \mathbb{R}^{|W|^2 \times |W|^2}$ with $Q(v, u) = -100, \forall u,v \in P$; $Q(v, u) = 0, \forall v\in P, \forall u \in N(v)$\;

\tcp{R-matrix init}
$R \in \mathbb{R}^{|W|^2 \times |W|^2}$ with $R(v, u) = -1, \forall u,v \in P$; $R(v, u) = 0,\forall v\in P\backslash \{v_0\}, \forall u \in N(v) \setminus \{v_0\}$, $R(v, v_0) = 100, \forall v \in N(v_0)$\;

\For{$m = 1$ to $M$}{
    Load $G_m = (V_m, E_m)$ with node positions and edges where Euclidean distance $\leq \mathcal{R}$\;

    \For{$k = 1$ to $K$}{
        Select random starting node $v \in V_m \setminus \{v_0\}$\;

        \While{$v \neq v_0$}{
            \eIf{random() $< \epsilon$}{
                \tcp{Exploration}
                Select random $u \in N^{V_m}(v)$;
            }{
                \tcp{Exploitation with random tie-breaking}
                Select $u = \arg\max\limits_{u' \in N^{V_m}(v)} Q(v, u')$;
            }
            Observe reward $r = R(v, u)$\;

            Find max Q at next state (random if multiple): 
            $\max_{index} = \arg\max\limits_{u' \in N^{V_m}(u)} Q(u, u')$ \;

            \tcp{Bellman update}
            Update $Q(v, u)$ using Eq. \ref{eq:bellman}\; 

            Set $v \gets u$\;
        }
    }
}
Save trained Q-matrix\;
\end{algorithm}

\begin{algorithm}[htbp]
\small
\caption{SPT Construction in Testing Phase}
\label{alg:qlearning_testing}
\KwIn{Trained Q-matrix, number of test graphs $T$}
\KwOut{SPT $T_t = (V, E_{T_t})$ for each test graph $G_t$, average routing accuracy}

\tcp{Note: Nodes mentioned in the algorithm refer to their locations, consisting of x and y coordinates.}
\For{$t = 1$ to $T$}{
    Load $G_t = (V, E)$ with nodes $v$ at positions $(x_v, y_v)$ and edges where Euclidean distance $\leq R$\;

    \tcp{Initialize directed tree} 
    $T_t = (V, E_{T_t})$, with $E_{T_t} = \emptyset$\;
    \tcp{Initialize predicted hop lengths list}
    $predicted\_layers = []$\;

    \For{each node $v \in V \setminus \{v_0\}$}{
        Initialize path $P = \{v\}$, current node $c = v$\;

        \While{$c \neq v_0$ and $|P| \leq |V|$}{
            \tcp{Exclude visited to avoid loops}
            Get neighbors $N^V(c) \setminus P = \{ u \in N^V(c) \mid u \notin P \}$;

            \If{$N(c) \setminus P = \emptyset$}{
                \tcp{No valid path}
                Break\; 
            }

            Compute scores: \\
            \tcp{$d(u,v_0)$ Euclidean distance between $u$ and sink $v_0$}
            $S(u) = Q(c, u) - d(u,v_0), \forall u \in N^V(c) \setminus P$\;
            $u = \arg\max\limits_{u \in N(c) \setminus P} S(u)$;

            $P \gets P \cup \{u\}$\;
            $E_{T_t} \gets E_{T_t} \cup \{(c, u)\}$\;
            $c \gets u$\;
        }
        $predicted\_layers[v] = |P| - 1$;
    }
    $predicted\_layers[v_0] = 0$\;
    Compute $label\_layers[v] = \min { |p| \mid p \text{ is a path from $v$ to $v_0$ in $G_t$} }, \forall v \in V$ using Dijkstra's Algorithm\;
    Compute accuracy: fraction of nodes where $predicted\_layers[v] = label\_layers[v]$\;
}
Compute average accuracy over all test graphs\;
\end{algorithm}

In the training phase, nodes iteratively update their Q-tables across $M = 5000$ network instances $G_m = (V_m, E_m)$, each with $K = 500000$ episodes, to learn robust routing policies. The reward function is defined in Eq. \ref{Eq: reward} incentivizing shortest paths by rewarding only sink-reaching transitions. In the testing phase, the trained Q-table is used to construct SPTs on $T = 100$ unseen networks. The Euclidean distance heuristic enhances path selection by prioritizing neighbors closer to the sink, enhancing efficiency while preserving loop-freedom. The distributed nature of the algorithm ensures scalability, as nodes operate independently, exchanging only $O(1)$ Q-value summaries per update.

\subsection{Theoretical Analysis}
\subsubsection{Convergence and Optimality}
The Q-learning algorithm converges to the optimal SPT under standard conditions for tabular RL in finite Markov decision processes (MDPs). The network is modeled as an MDP with states as nodes, actions as next-hop selections from $N^{V_m}(v)$, and rewards favoring the sink. With sufficient exploration ($\epsilon > 0$), the Q-values satisfy the Bellman optimality equation at convergence:

\begin{equation}
Q^*(v, u) = r(v, u) + \gamma \max_{u' \in N^{V_m}(u)} Q^*(u, u')
\end{equation}

Since rewards are 100 at the sink and 0 otherwise, and $\gamma < 1$, Q-values decrease with hop distance, ensuring the policy $\pi(v) = \arg\max\limits_u Q(v, u)$ selects neighbors on shortest paths. Proof of convergence follows \cite{watkins1992q} given infinite state-action visits.

\subsubsection{Loop-Freedom}

The converged policy is loop-free because loops yield no reward (0 cumulative) and are dominated by paths to the sink (positive reward). Exploration breaks potential loops, as suboptimal Q-values for looping actions are lower than those for sink-reaching paths.

\subsubsection{Complexity}
\textbf{Computational}: Each update is $O(|N^{V_m}(v)|)$, with $K \times M$ episodes; for typical IoT degrees ($|N^{V_m}(v)| \approx 10$), training is efficient. \textbf{Communication}: Local Q-summary exchanges; no global flooding, $O(1)$ per update. Compared to Dijkstra's ($O(N^2$)) centralized cost. This is highly scalable for distributed IoT.

\label{4_Proposed_Approach}

\section{Performance Evaluation}

To validate the proposed Q-learning-based approach for SPT construction in IoT sensor networks, we conducted extensive simulations across various network sizes and configurations. This section describes the experimental setup, training and testing protocols, performance metrics, and results, including comparisons with the optimal SPTs computed by Dijkstra’s algorithm. We also present visualizations to illustrate the effectiveness of the learned routing trees.

\subsection{Simulation Setup}

We simulated wireless sensor networks in a 100x100 unit\(^2\) area, with the following parameters:
\begin{itemize}
    \item \textbf{Network Size}: $N = 100, 200, 300, 400, 500$ nodes, randomly deployed with uniform distribution.
    \item \textbf{Communication Range}: $R = 20$ units, where a link $(u, v) \in E$ exists if the Euclidean distance between nodes $u$ and $v$ is at most $R$.
    \item \textbf{Sink Location}: The sink node $v_0$ is fixed at coordinates $(50, 50)$, representing a centralized data collection point.
\end{itemize}

The resulting graphs are random geometric graphs, a standard model for ad hoc sensor networks. Network connectivity was ensured by verifying that all nodes have at least one multi-hop path to the sink. Node positions were assigned integer coordinates $(x_v, y_v)$, and edges were added based on the communication range.

\subsection{Training Protocol}

For each network size $N$, we trained the Q-learning algorithm on 5,000 randomly generated network instances to build robust Q-table. The training parameters were:


\begin{itemize}
    \item \textbf{Episodes}: 500,000 episodes per graph, initiating at a random non-sink node and continuing until the sink node is reached.
    \item \textbf{Learning Rate}: $\alpha = 0.9$, set to favor recent updates while preserving prior knowledge.
    \item \textbf{Discount Factor}: $\gamma = 0.9$, prioritizing shorter paths via discounted future rewards.
    \item \textbf{Exploration Probability}: $\epsilon = 0.5$, promoting exploration of diverse routing paths.
\end{itemize}

The Q-table was initialized with $Q(v, u) = -100$ for invalid actions ($u \notin N(v)$) and $Q(v, u) = 0$ for valid neighbors $u \in N(v)$. The reward matrix was set with $R(v, v_0) = 100$ for neighbors of the sink, $R(v, u) = 0$ for other valid neighbors, and $R(v, u) = -1$ for invalid actions. Training was conducted across multiple graphs to capture general routing patterns, with Q-values updated using the Bellman equation as described in Section \ref{4_Proposed_Approach}.

\subsection{Testing Methodology}

After training, we evaluated the learned Q-table on 100 unseen network instances for each network size. Two testing scenarios were considered:
\begin{enumerate}
    \item \textbf{Same-Size Testing}: Q-table trained on a specific network size (e.g., 300 nodes) were tested on new networks of the same size.
    \item \textbf{Cross-Size Testing}: Q-table trained on larger networks (300, 400, or 500 nodes) were applied to networks of different sizes (100, 200, 300, 400, 500 nodes) to assess generalization.
\end{enumerate}

For each test graph $G_t = (V, E)$, the SPT was constructed by selecting the next-hop node for each $v \in V \setminus \{v_0\}$ using a combined score: $S(u) = Q(v, u) - d(\text{pos}(u), \text{pos}(v_0))$, where $d(\text{pos}(u), \text{pos}(v_0))$ is the Euclidean distance of nodes $u$ and $v_0$, $\text{pos}(u)$ denotes the coordinates of node $u$. This heuristic prioritizes neighbors closer to the sink, enhancing path selection efficiency. The resulting directed edges formed the SPT $T_t = (V, E_{T_t})$. To avoid loops, nodes already visited in a path were excluded from consideration.

The ground truth SPT was calculated using Dijkstra’s algorithm (implemented as a breadth-first search for unit-weight edges), serving as the benchmark for optimality. The path lengths from each node to the sink in the learned SPT were compared to those of the optimal SPT to evaluate precision.

\subsection{Performance Metrics}

We used \textbf{routing accuracy} metric to assess performance. The routing accuracy is the percentage of nodes whose path length to the sink in the learned SPT matches the shortest path length in the optimal SPT, computed as:

\begin{equation}
    \text{Accuracy} = \frac{|\{ v \in V \mid d_{T_t}(v, v_0) = d_G(v, v_0) \}|}{|V|},
\label{eq:acc}
\end{equation}

where $d_{T_t}(v, v_0)$ is the path length in the learned tree $T_t$, and $d_G(v, v_0)$ is the shortest path length in $G_t$. An accuracy of 100\% indicates the learned tree is identical to the optimal SPT.

\subsection{Results}

\subsubsection{Same-Size Network Results}

Table~\ref{tab:same_size_accuracy} presents the routing accuracy for same-size network testing, averaged over 100 test networks per size.

\begin{table}[htbp]
\centering
\caption{Routing Accuracy for Same-Size Network Testing}
\resizebox{0.7\columnwidth}{!}{
\begin{tabular}{|c|c|}
\hline
\textbf{Network Size ($N$)} & \textbf{Accuracy (\%)} \\ \hline
100 & 82.15 \\ \hline
200 & 98.33 \\ \hline
300 & 99.46 \\ \hline
400 & 99.72 \\ \hline
500 & \textbf{99.80} \\ \hline
\end{tabular}
}
\label{tab:same_size_accuracy}
\end{table}

The results demonstrate high accuracy, exceeding 98\% for networks with 200 or more nodes. The lower accuracy for 100-node networks (82.15\%) is attributed to boundary effects and fewer hops, which reduce the differentiation between optimal and suboptimal paths, requiring more exploration. For larger networks, increased connectivity and hop counts facilitate clearer identification of optimal routes, leading to near-optimal SPTs.

\subsubsection{Cross-Size Network Results}

Table~\ref{tab:cross_size_accuracy} shows the accuracy when applying Q-table trained on larger networks (300, 400, 500 nodes) to different network sizes, averaged over 100 test networks.

\begin{table}[htbp]
\centering
\caption{Routing Accuracy for Cross-Size Network Testing}
\begin{adjustbox}{max width=\columnwidth}
\begin{tabular}{|c|c|c|c|}
\hline
\textbf{Test Size} & \textbf{300-Node (\%)} & \textbf{400-Node (\%)} & \textbf{500-Node (\%)} \\ \hline
100 & 94.87 & 95.29 & 95.30 \\ \hline
200 & 99.00 & 99.32 & 99.32 \\ \hline
300 & 99.46 & 99.58 & 99.58 \\ \hline
400 & 99.67 & 99.72 & 99.72 \\ \hline
500 & 99.74 & \textbf{99.80} & \textbf{99.80} \\ \hline
\end{tabular}
\end{adjustbox}
\label{tab:cross_size_accuracy}
\end{table}

The Q-table trained on larger networks generalize effectively to smaller networks. For instance, the 500-node Q-table achieves 95.30\% accuracy on 100-node networks, comparable to direct training on those sizes. This suggests that the learned policies capture general routing principles applicable to similar topology distributions, reducing the need for retraining in varied network sizes.

\subsubsection{Visualization}

\begin{figure}[htbp]
\centering
\begin{tabular}{cc}
\includegraphics[width=0.45\columnwidth]{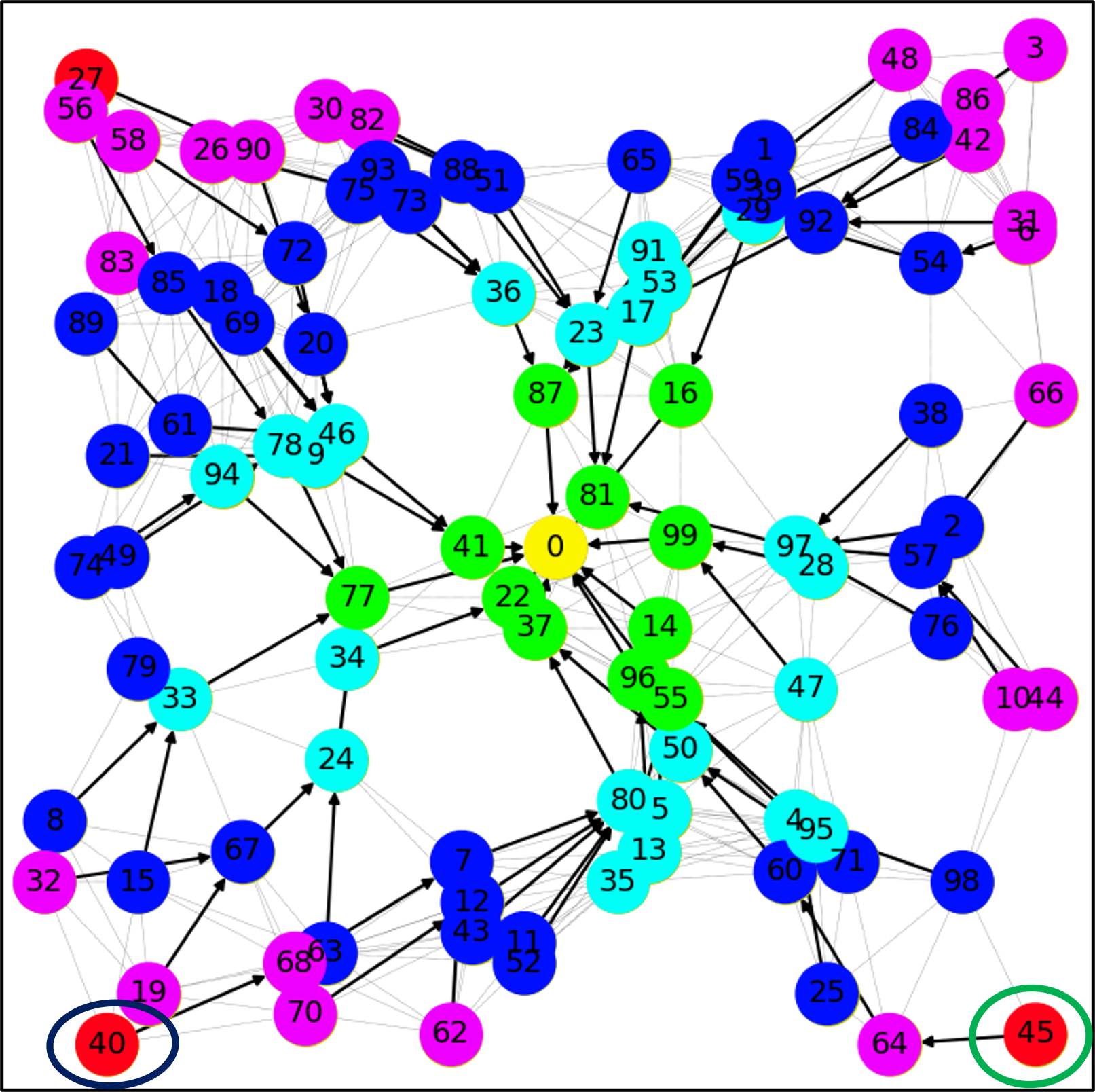} & \includegraphics[width=0.45\columnwidth]{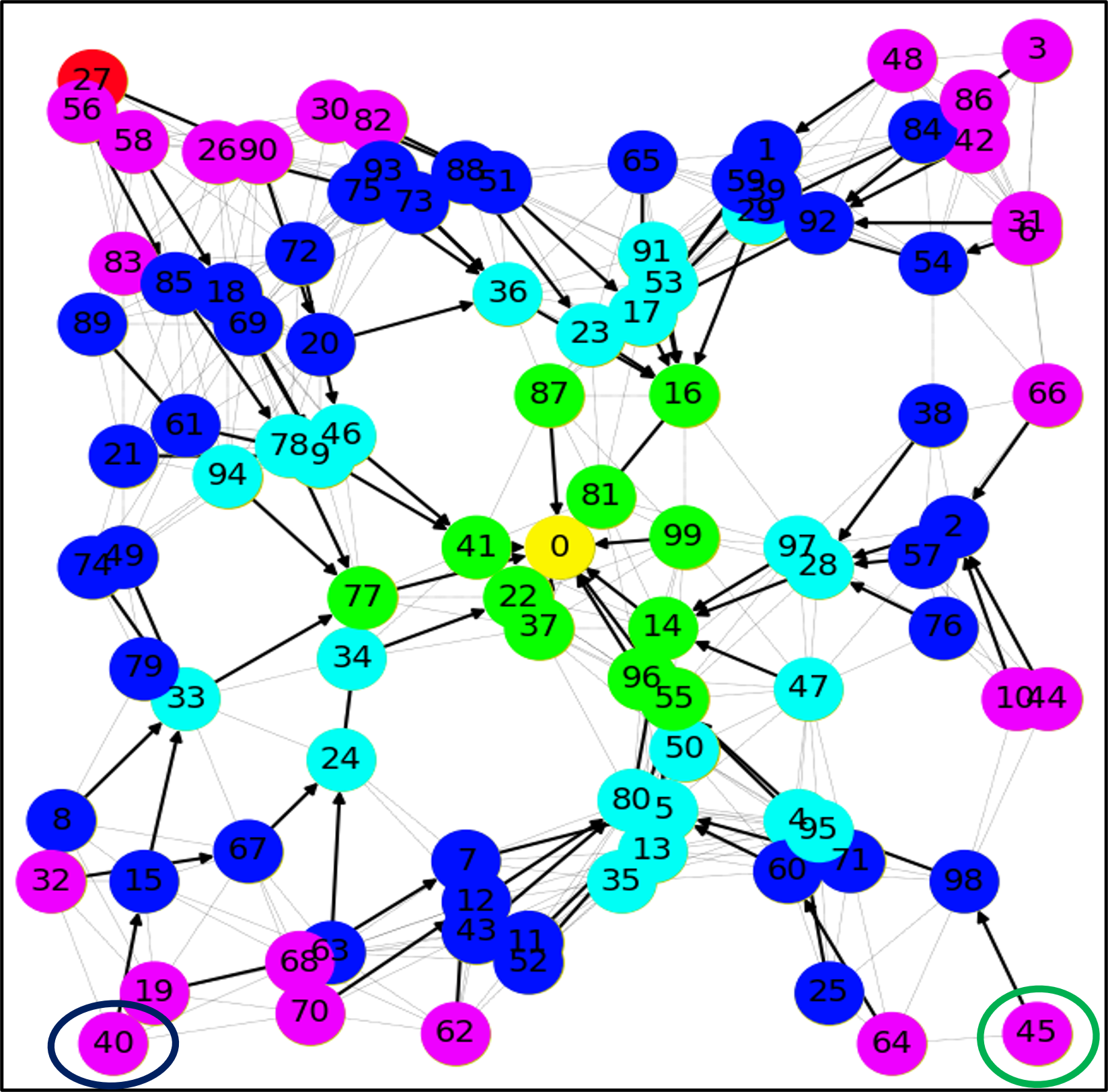} \\
(a) Q-Learning & (b) Dijkstra’s Algorithm \\
\end{tabular}
\caption{SPT for 100-Node Network}
\label{fig:spt_100}
\end{figure}

\begin{figure}[htbp]
\centering

\begin{tabular}{cc}
\includegraphics[width=0.45\columnwidth]{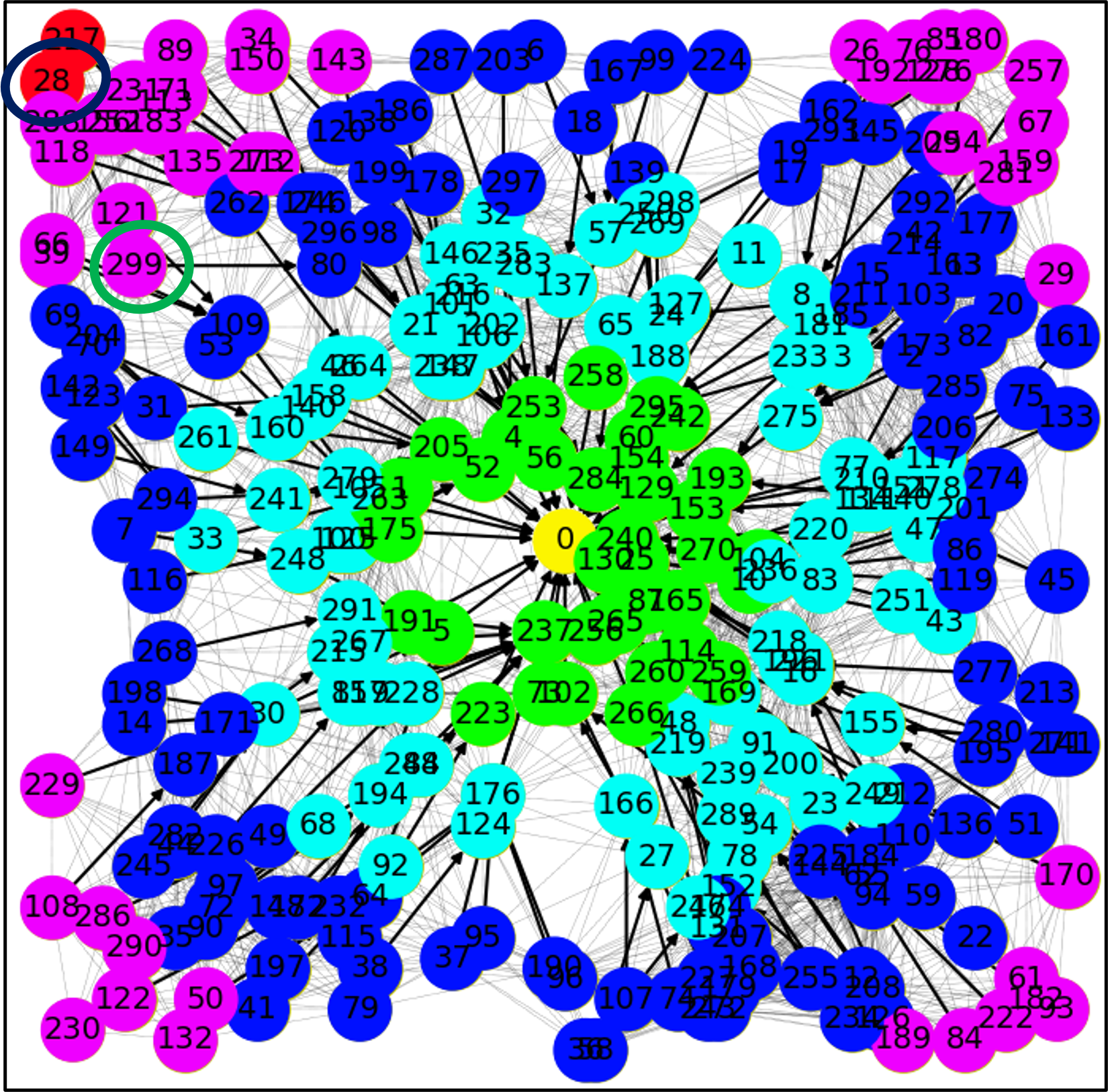} & \includegraphics[width=0.45\columnwidth]{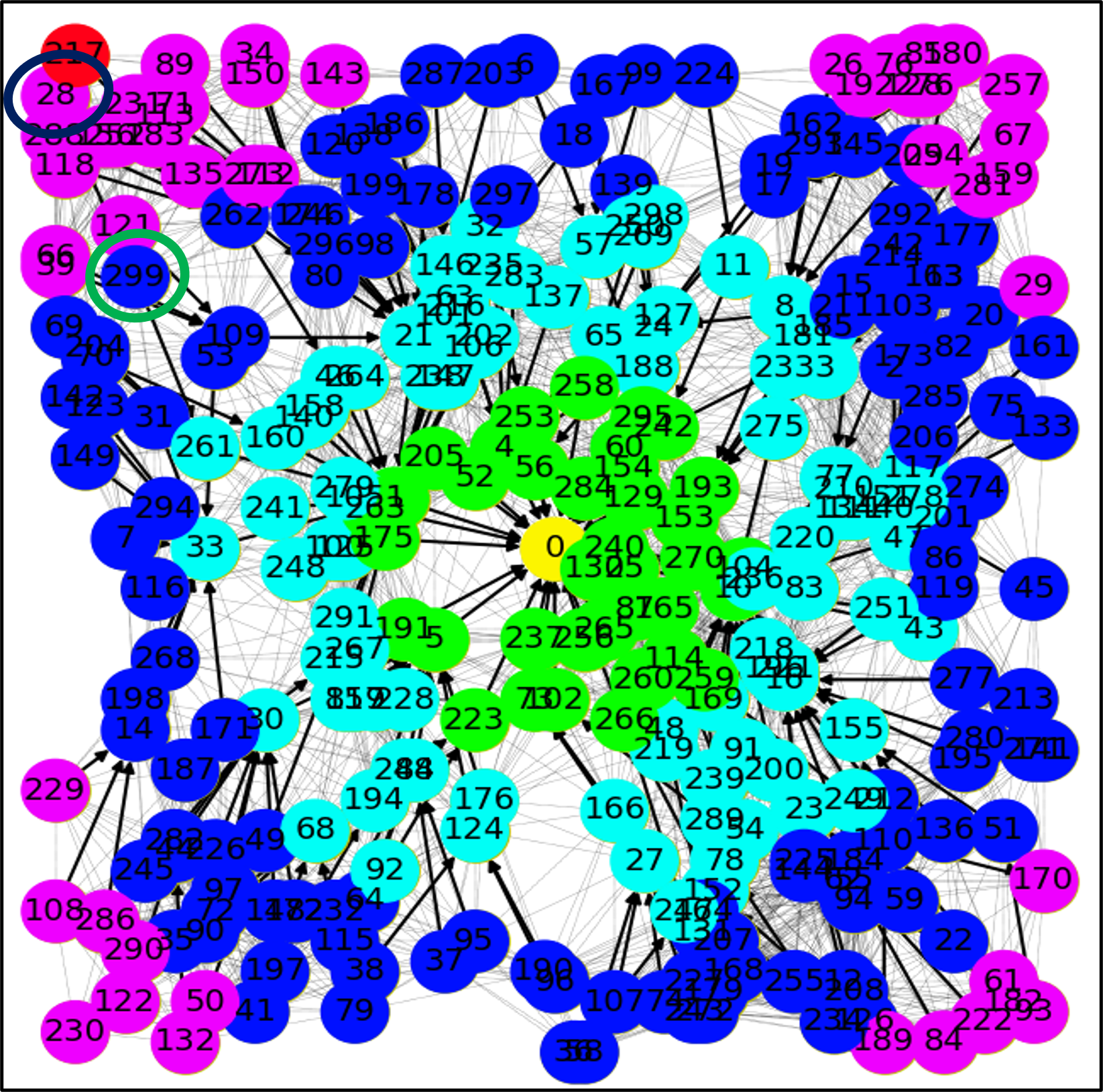} \\
(a) Q-Learning & (b) Dijkstra’s Algorithm \\
\end{tabular}
\label{fig:spt_300}
\caption{SPT for 300-Node Network}
\end{figure}

\begin{figure}[htbp]
\centering
\begin{tabular}{cc}
\includegraphics[width=0.45\columnwidth]{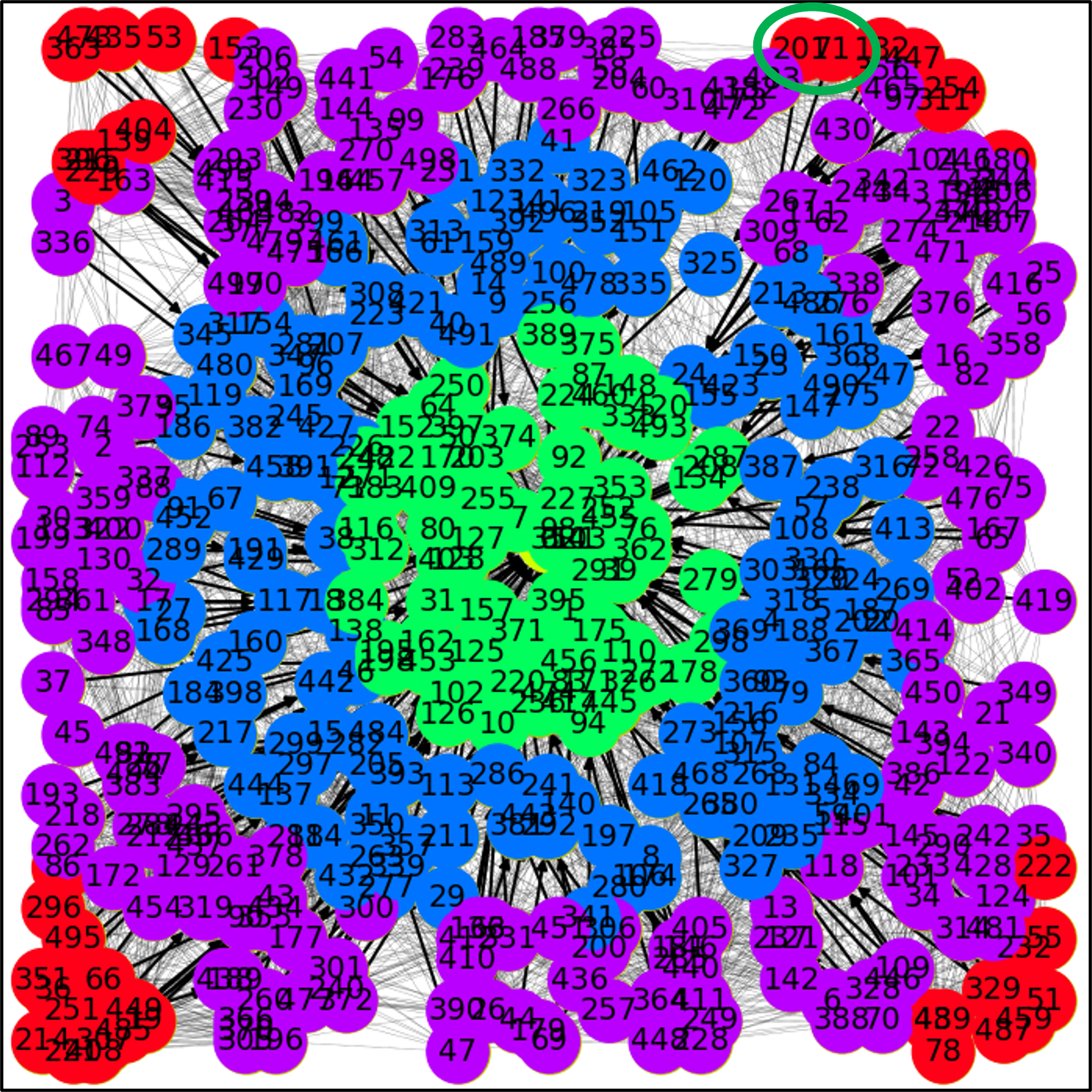} & \includegraphics[width=0.45\columnwidth]{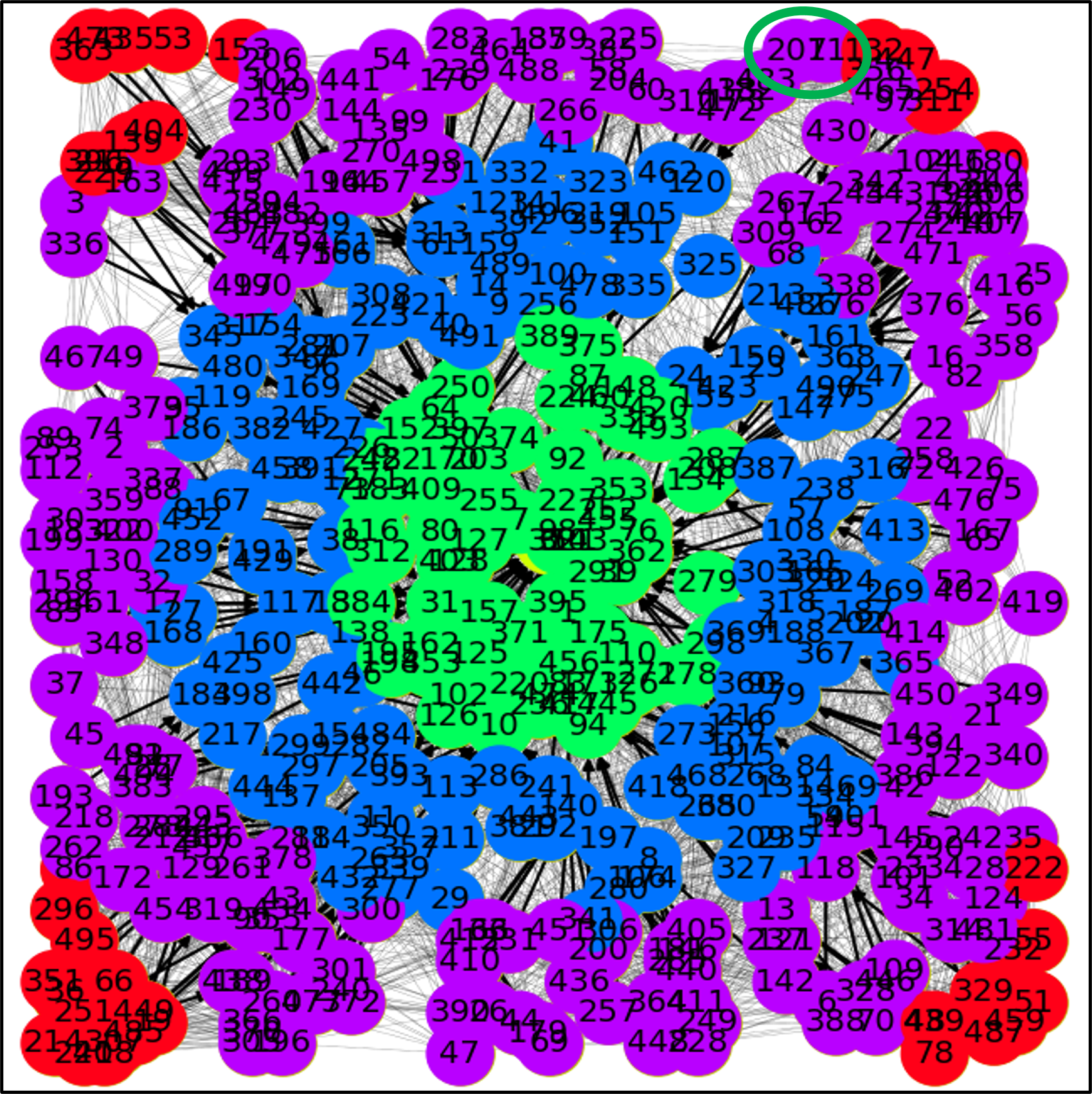} \\
(a) Q-Learning & (b) Dijkstra’s Algorithm \\
\end{tabular}
\caption{SPT for 500-Node Network}
\label{fig:spt_500}
\end{figure}

Figs.~\ref{fig:spt_100} to~\ref{fig:spt_500} visualize the SPTs constructed by Q-learning compared to those by Dijkstra’s algorithm for networks of 100, 300, and 500 nodes. The Q-learning SPTs are plotted with nodes colored by their hop distance to the sink, overlaid on the undirected graph $G_t$ (shown with dotted edges). The Dijkstra SPTs serve as the ground truth.

For 100-node networks, the Q-learning SPT shows minor deviations at the periphery due to exploration challenges, reflected in the lower accuracy (82.15\%). For 300 and 500 nodes, the learned SPTs are nearly identical to the optimal SPTs, with accuracy exceeding 98\%, confirming the algorithm’s effectiveness in larger networks.

\subsection{Discussion}

The experimental results validate the proposed Q-learning approach, achieving near-optimal routing accuracy (over 99\% for $N \geq 300$) and scalability across network sizes. The Euclidean distance heuristic in testing enhances path selection, prioritizing sink-proximate neighbors for efficient, loop-free SPTs. Cross-size testing highlights generalization, with Q-table from 500-node training performing comparably on smaller networks, enabling pre-training on representative topologies to minimize deployment overhead in IoT systems. Visualizations in Figs.~\ref{fig:spt_100} to~\ref{fig:spt_500} show learned SPTs closely matching Dijkstra's optima, confirming robustness even in smaller networks.

Minor errors in smaller networks (e.g., 100 nodes) yield paths 1-2 hops longer, with negligible impact on energy and latency compared to centralized overheads like Dijkstra's or flooding in RPL. High accuracy in larger networks mitigates these, leveraging connectivity for optimal differentiation. Compared to global-knowledge methods, our distributed framework suits IoT's limited resources. Future work includes baselines like greedy routing, dynamic evaluations (e.g., failures), and deep RL extensions for larger scales.

\label{5_Experiment_Results}

\section{Conclusion}
This paper introduces a distributed Q-learning framework for constructing SPTs in IoT sensor networks, allowing nodes to learn optimal next-hop decisions with local information alone. Eliminating centralized control, it suits resource-constrained environments. Theoretical analysis verifies convergence to an optimal SPT with low overhead. Simulations on 100 to 500 nodes show near-optimal accuracy (over 99\% for $N \geq 300$), with minor deviations (1--2 extra hops) in smaller networks having negligible energy and latency impact.

Outperforming centralized (e.g., Dijkstra's) and flooding-based methods, our approach provides distributed, adaptive routing and generalizes across network sizes for efficient deployment. Limitations include training overhead and parameter sensitivity in dynamic settings. Future directions involve deep RL for larger scales, integrating link reliability and energy-aware routing, and real IoT hardware validation to handle packet loss and heterogeneity. This work advances autonomous optimization in IoT, inspiring further RL-network protocol research for resilient systems.

\label{6_Conclusion}



\printbibliography

\end{document}